\documentclass[twocolumn,prb,showpacs]{revtex4}

\usepackage{epsfig,amsmath,amssymb,color,graphicx}
\begin{document}

%\title{The density matrix renormalization group method as a solver for cluster perturbation theory: origin of the pseudogap the 2D Hubbard model}
\title{Spectral function of the 2D Hubbard model: a density matrix renormalization group plus cluster perturbation theory study}
\author{Chun Yang}
\affiliation{Department of Physics, Northeastern University, Boston, Massachusetts 02115, USA}
\author{Adrian E. Feiguin}
\affiliation{Department of Physics, Northeastern University, Boston, Massachusetts 02115, USA}

\date{\today}
\begin{abstract}
We study the spectral function of the 2D Hubbard model using cluster perturbation theory, and the density matrix renormalization group as a cluster solver. We reconstruct the two-dimensional dispersion at, and away from half-filling using $2\times L$ ladders, with $L$ up to 80 sites, yielding results with unprecedented resolution in excellent agreement with quantum Monte Carlo. The main features of the spectrum can be described with a mean-field dispersion, while kinks and pseudogap traced back to scattering between spin and charge degrees of freedom.
\end{abstract}
\maketitle

\section{Introduction}

 Mott insulators defy conventional paradigms, since the rigid band picture behind the physics of semiconductors does not apply: in strongly interacting systems, the bands change with doping, giving rise to a complex phenomenology that includes hole pockets, Fermi arcs and kinks \cite{Damascelli2003,Ronning2005,Graf2007,Meng2009}.
The spectral properties near the Mott transition in the Hubbard model have been studied extensively by a number of computational techniques \cite{Dagotto1991e,Ohta1992,Dagotto1992,Meinders1993,Eskes1994,Preuss1994a,Bulut1994,Preuss1995,Moreo1995c,Georges1996,Preuss1997,Groeber2000,Huscroft2001,Senechal2000,Maier2002a,Senechal2002,Stanescu2003,Dahnken2004,Aichhorn2005,Stanescu2006,Aichhorn2006,Macridin2006,Tremblay2006,Kyung2006,Stanescu2006a,Sakai2009,Liebsch2010,Sakai2010,Kohno2010,Eder2011,Kohno2012,Wang2015a,Kung2015} and results indicate the emergence of excitations in the Mott gap at finite doping. The ``leaking'' of spectral weight into the gap has been explained a while ago by a seminal work by Eskes {\it et al} \cite{Eskes1991}, and reviewed in Ref.\onlinecite{PhillipsRMP2010}.

Previous numerical studies using cluster perturbation theory (CPT) \cite{Kohno2012} indicate the survival of one-dimensional aspects in the spectrum of the fully two-dimensional system, and suggest that some of the features observed in the experiments, such as kink or waterfalls \cite{Damascelli2003} could be attributed to spin-charge separation and traced back adiabatically to spinon and holon dispersion in one-dimensional chains.

In one-dimensional (1D) systems, the Fermi-liquid picture breaks down: the natural excitations are described by Luttinger liquid theory\cite{GiamarchiBook,Gogolin,Haldane1981} as collective bosonic modes carrying spin and charge, with each degree of freedom being characterized by a different energy scale. 
Even though spin-charge separation is intrinsically a manifestation of
1D physics, the possibility of its presence in two-dimensions
(2D), or quasi-2D systems has been extensively debated, particularly
within the context of high-temperature superconductivity\cite{Anderson2000}.
Some numerical studies in this
direction, looking at 2,3 and 4-leg $t-J$ ladders, indicate the presence
of spinon and holon excitations \cite{Poilblanc1995,Haas1996a,Rice1997,Martins2000,Brunner2001}.
Whether spin-charge separation, or electron-phonon interactions are responsible for the unexpected spectral features such as kinks, and 'waterfalls' 
in cuprates, is still open to interpretation and a topic of great debate. 

Since CPT relies on the solution of small clusters, it cannot describe long-range order. These shortcomings can be overcome by using an extension of the method called the variational cluster approximation (VCA) or also referred-to as VCPT. The VCA extends the previous ideas by incorporating additional ingredients, such as external fields, or even additional cluster sites, and introducing a variational principle to self-consistently determine the optimal symmetry-breaking fields \cite{Potthoff2003,Potthoff2003a,Potthoff2003b,Dahnken2004,Aichhorn2006,Potthoff2006}.
The variational principle is derived from a general framework called self-enegy functional approach that has the power to unify several cluster methods, including cluster (or cellular) dynamical mean field theory (C-DMFT) \cite{Kotliar2001} and dynamical cluster approximation (DCA),\cite{Hettler1998,Hettler2000} under the same mathematical structure \cite{Maier2005}.

In this work, we apply the time-dependent density matrix renormalization group method (tDMRG) \cite{White2004a,Daley2004,Feiguin2005,vietri} as a solver for CPT, and we use it to study the spectral function of the 2D Hubbard model with unprecedented resolution at, and away from half-filling.
The tDMRG allows us to couple clusters that are already infinite (very large) in one spatial dimension, representing a tremendous advance over traditional calculations with small clusters, with typically 12-16 sites. 
%In addition, we are able to identify behavior that can be related to the onset of quasi-long-range order, bridging results from CPT, VCA, DCA, and quantum Monte Carlo in a coherent picture for understanding the pseudogap formation in the two-dimensional Hubbard model. 

 In Section II we introduce the methods, in Section III we describe and analyze the results, and we close with a discussion.

\section{Methods}
Cluster perturbation theory (CPT) is an technique that applies to problems with local interactions, such as the Hubbard model\cite{Gross1993,Senechal2000,Senechal2002}.
It provides an approximation to the single particle Green's function of the problem in the thermodynamic limit by coupling clusters of small size in a variation of strong coupling perturbation theory.
The main idea consists of dividing the lattice into small clusters which can be diagonalized exactly, and coupling them together to reconstruct the original system. The single particle Green's function for the thermodynamic limit is constructed by solving a simple Dyson's equation:
\begin{equation}
\bf{G}^{-1} = \bf{G'}^{-1}-\bf{T}, \label{dyson}
\end{equation}
where the bold symbols represent matrices: $\bf{G}$ is the Green's function we seek, $\bf{G'}$ is the Green's function in the cluster, and $\bf{T}$ is a hopping matrix connecting the clusters. %The Green's functions for the cluster are typically obtained using exact diagonalization or the Lanczos algorithm.
In the following we assume that the symbol ${G}$ refers to {\it retarded} Green's functions.

\begin{figure}%[ht]
 \includegraphics[width=0.48\textwidth, angle=0]{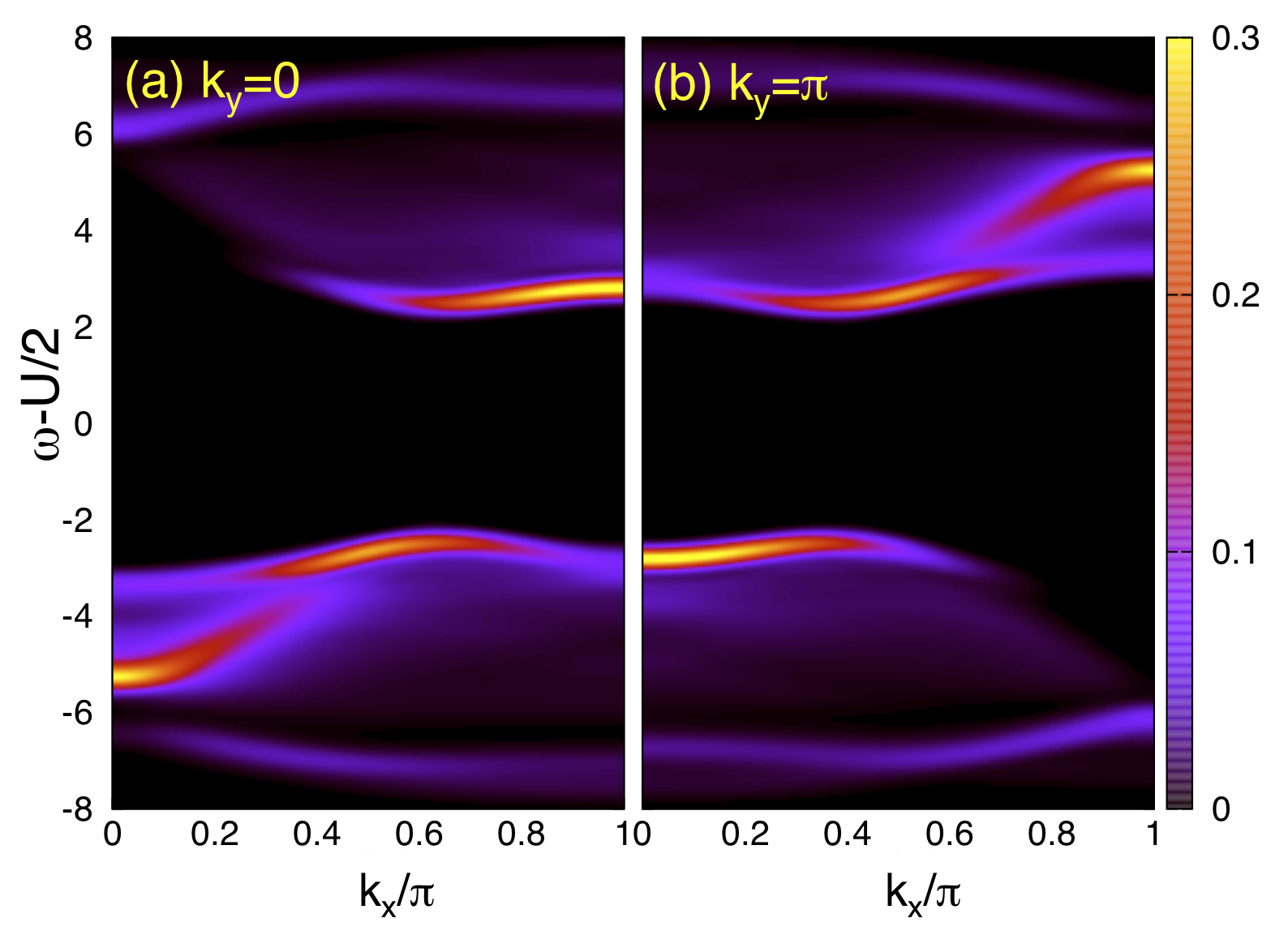}
\caption{
Spectral function of a Hubbard ladder with $L=80$ and $U/t=8$, at half-filling, obtained with tDMRG. Panels (a) and (b) show the symmetric and anti-symmetric sectors, respectively, which are related by particle-hole symmetry.
}
\label{fig:ladder}
\end{figure}

In this work, our cluster consists of a $2 \times L$ ladder, and the model is given by the usual Hubbard Hamiltonian:

\begin{eqnarray}
H = & - & t\sum_{i,\lambda,\sigma}\left(c^\dagger_{i,\lambda\sigma} c^{\phantom{\dagger}}_{i+1,\lambda\sigma}+\mathrm{h.c.}\right) + \nonumber\\
& - & t\sum_{i,\sigma}\left(c^\dagger_{i2\sigma} c^{\phantom{\dagger}}_{i1\sigma}+\mathrm{h.c.}\right) + U \sum_{i,\lambda} n_{i,\lambda\uparrow}n_{i,\lambda\downarrow},
\end{eqnarray}
where the operator  $c^\dagger_{i\lambda\sigma}$ creates an electron on rung $i$ and leg $\lambda=1,2$ with spin $\sigma$, $n_{i\lambda\sigma}$ is the electron number operator, and $t$ and $U$ parametrize the hopping and Coulomb repulsion, respectively.
In the following we assume periodic boundary conditions in the leg direction, and we will address the finite size effects in the technical discussion below.

Since the cluster possesses translational invariant along the leg direction $x$, we can readily Fourier transform our Green's functions as:
\[
G'_{\lambda \lambda'}(k_x)=\sum_n e^{ik_x na} G'_{\lambda \lambda'}(x),
\]
where $a$ is the lattice spacing, $x=na$, and we have omitted the spin index, since our problem is also invariant under a spin inversion. This expression defines a Green's function in a hybrid representation, since the leg index $\lambda$ still represents a real space coordinate. However, Eq.(\ref{dyson}) is diagonal in $k_x$, meaning that $G$ is a $2 \times 2$ matrix for each value of $k_x$, which is exactly equivalent to solving the CPT equations for a $2$-site cluster: 
\[
G^{-1}_{\lambda,\lambda'}(k_x, Q, \omega) = G'^{-1}_{\lambda,\lambda'}(k_x, \omega) - T_{\lambda,\lambda'}(Q),
\]
with 
\[
T_{\lambda,\lambda'}(Q) = -t \left[e^{iQ} \delta_{\lambda,2}\delta_{\lambda',1} + e^{-iQ}\delta_{\lambda,1}\delta_{\lambda',2} \right]
\]
and $Q=2k_y$ introducing the dependence on $k_y$.
By restoring the quasi-translational invariance, we obtain the CPT Green's function as:
\begin{equation}
G^{CPT} (k_x,k_y,\omega) = \frac{1}{2}\sum_{\lambda,\lambda'=1}^2 e^{-ik_y(\lambda-\lambda')}G_{\lambda,\lambda'}(k_x,2k_y,\omega).
\label{gcpt}
\end{equation}

\begin{figure}%[ht]
 \includegraphics[width=0.48\textwidth, angle=0]{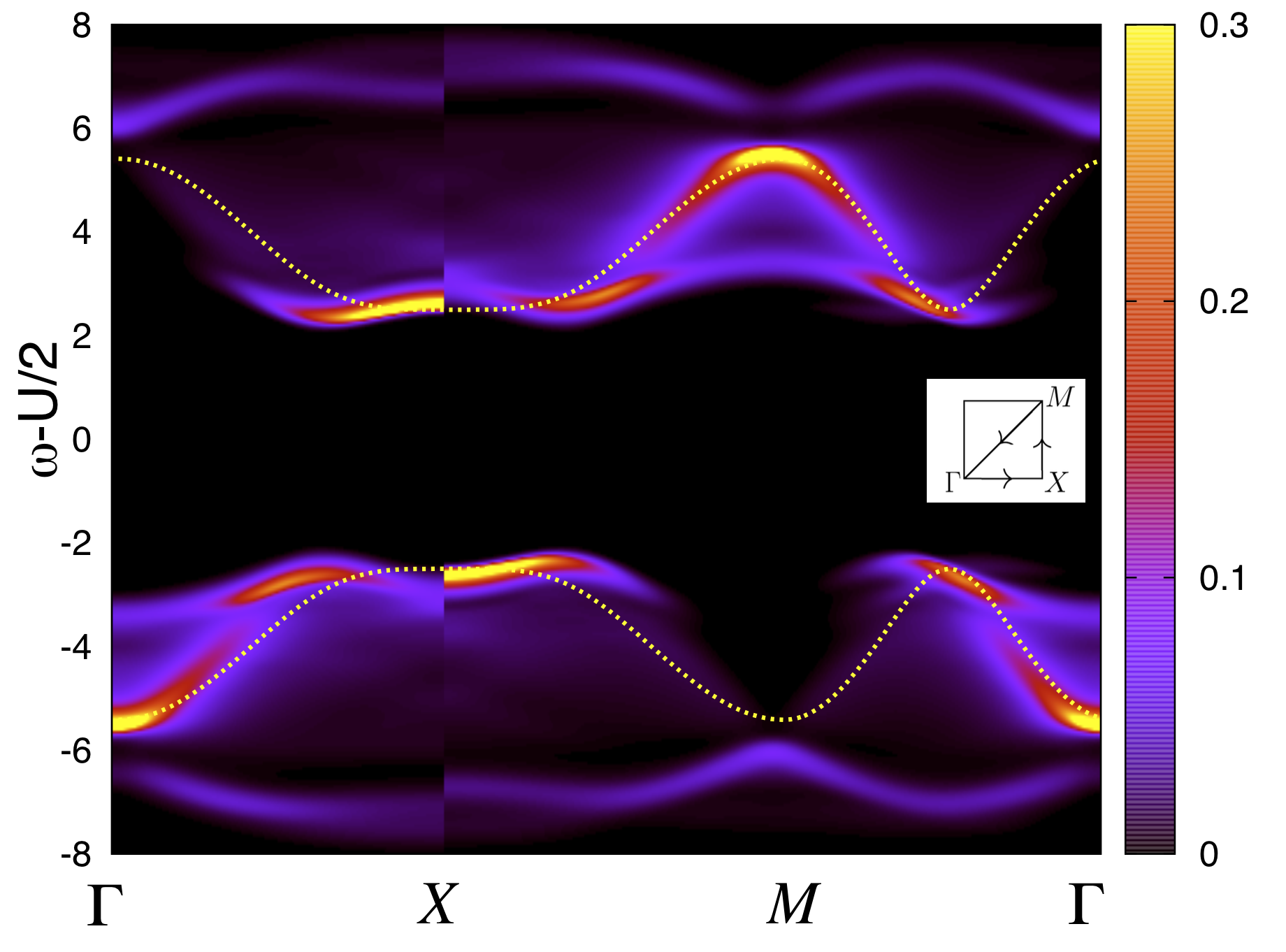}
\caption{
CPT spectral function of the $U=8$ 2D Hubbard model at half-filling obtained using a $2\times 80$ ladder as a cluster, and tDMRG as a solver. The dashed line shows the Hartree-Fock dispersion.
}
\label{fig:half-filling}
\end{figure}

In addition, by symmetry we obtain $G_{11}=G_{22}$ and $G_{21}=G_{12}$, which reduces the number of required simulations.
These equations are clearly very simple, and the main challenge consists of calculating the Green's functions with tDMRG, which can be readily done using well established methods, extensively described in the literature\cite{Feiguin2005}, and reviewed in Ref.\onlinecite{vietri}. 
The tDMRG method yields the single particle correlation function in real space and time, and the Green's functions are obtained by Fourier transforming the results to momentum and frequency. 
The most subtle aspect of the calculation concerns the use of open boundary conditions along the $x$ direction. 
As discussed in Refs.\onlinecite{White2004a,Feiguin2005,Pereira2009}, the finite size effects introduced by the boundaries can be controlled in two ways: by convolving the Fourier transform to momentum space with a smooth window that vanishes at the boundaries, and by limiting the simulation time to prevent reflections at the two ends of the ladder. In addition, to avoid artifacts such as ``ringing'' in the Fourier transform to frequency, we also convolve the results with a Hann window along the time direction. This has the effect of introducing an artificial broadening in the spectral function that is inversely proportional to the width of the time window.
 Long simulation times would reduce the broadening in frequency, with the price of introducing ringing. These features are amplified when the matrix is inverted and plugged into the CPT equation, introducing instabilities that result, for instance, in negative values of the spectral function. 
Therefore, our simulation times (and Hann window width) are relatively short $t_\mathrm{max} \sim 15$ in units of the hopping, and make the use of linear prediction methods to extrapolate in time \cite{White2008} completely unnecessary. 

\section{Results}

We have simulated a $2\times80$ Hubbard ladder with 600 DMRG states, and using a time window of width $\Delta t=15$, a time step $\delta t=0.02$, and a third order Suzuki-Trotter decomposition of the evolution operator (In the following, we take $t=1$ as our unit of energy).
In Figs.\ref{fig:ladder}(a) and (b) we show results for the bare spectral function of the ladder (before CPT), at half-flling and for $U/t=8$, as a function of $k_x$, and for the symmetric and antisymmetric sectors, represented by $k_y=0, \pi$, respectively:
\[
G'(k_x,k_y=0,\pi,\omega)= G'_{11}(k_x,\omega) \pm G'_{12}(k_x,\omega),
\]
where the $\pm$ signs correspond to the two values of $k_y$.
Interestingly, the truncation errors are very small, of the order of $10^{-7}$, which can be explained by noticing that the cluster is gapped in both the charge and spin sectors. 
Curiously, and to the best of our knowledge, there are no results with DMRG for this ladder system in the literature, probably stemming from previous observations that {\it dynamical} DMRG\cite{Hallberg1995,Kuhner1999,Jeckelmann2002} is computationally very expensive in this geometry, and only recently it has been applied to $t-J$ ladders \cite{Kohno2015}.

Even though ladders are quasi-one-dimensional systems with spin-charge separation and Luther-Emery behavior \cite{Luther1974,Balents1996}, the sharp features observed in chains, such as shadow and spinon bands, are washed out and less discernible, with most of the spectral weight concentrated in the holon bands. The spinon bands in the lower Hubbard band (LHB) for $k_y=0$ shows a tendency to merge with the holon band and form a single quasi-particle dispersion, as one would expect from a Fermi liquid. 
The dispersion presents a  ``waterfall'' that resembles a discontinuity in the dispersion at $k_x=\pi/2$, and could be attributed to a mixing between the charge and spinon modes. The upper Hubbard band (UHB) displays a sharp spinon-like dispersion centered at $k_x=\pi$ with very small band width. These features are reversed for $k_y=\pi$: due to particle-hole symmetry the bands are reflected about the Fermi energy and shifted in $k_x$ by $\pi$. 

In Fig.\ref{fig:half-filling}(a), we present the tDMRG+CPT spectral function of the 2D Hubbard model at half-filling with $U/t=8$ along the $\Gamma\rightarrow X \rightarrow M$ path in the Brillouin zone (BZ).
 The CPT equations along the $X \rightarrow M$ line will produce a mixture of $G'_{11}(\pi,\omega)$ and $G'_{12}(\pi,\omega)$. The small cluster size in the transverse direction yields very limited resolution along this line. However, in a rotational invariant lattice, they should be identical to the results for the $k_y=\pi$ boundary of the Brillouin zone, which can be obtained with very high resolution. For this reason, we have plotted the CPT spectrum for the $k_y=\pi$ along the $X\rightarrow M$ segment, with the price of introducing an artificial discontinuity at the $X$ point.

The spectrum shows and uncanny resemblance to the ladder's, albeit a weak renormalization. %, indicating that the cluster already contains information about the 2D physics.
As explained in Ref.\onlinecite{Kohno2012}, the CPT introduces a shift of spectral weight at high energies while keeping the spectral weight near the Fermi level almost unaffected which makes the holon-like bands sharper, and the spinon-like bands weaker, yielding a  dispersion that resembles that of quasi-particles. The spinon features remains as an incoherent background {\it at low energies}, while preserving the ``waterfall''  at $(\pi/2,\pi/2)$.

Following Ref.\onlinecite{Grober2000}, the quasi-particle dispersion can be fitted by a mean-field (Hartree-Fock) dispersion assumming a N\'eel antiferromagnetic order\cite{Eder2000,Grober2000} (AFM), given by the two bands
\[
E_{\pm}({\bf k})=\pm \sqrt{[-2\tilde{t}(\cos{k_x}+\cos{k_y})]^2+\tilde{\Delta}^2},
\]
as shown by the dashed line in the figure, where we take the gap $\tilde{\Delta}$ and $\tilde{t}$ as a free fitting parameters.
This indicates that, despite its low dimensionality, the ladder cluster already introduces features in the spectrum that contain information about the onset of AFM order. Moreover, the spectral function displays a remarkable agreement with the quantum Monte Carlo (QMC) results from Refs.\onlinecite{Bulut1994,Preuss1995,Preuss1997,Grober2000} but with much better resolution. In particular, we observe similar features such as the flat dispersion in the UHB and LHB centered at the $(\pi,0)$ point, and the weak spinon-like incoherent background at low energies. 
The high energy ``bands'' observed in QMC can be associated to the shadow bands in the ladder dispersion, echoes of one-dimensional physics. Ramarkably, these same features are also obtained using square clusters with CPT \cite{Kohno2012}, and VCPT \cite{Dahnken2004,Aichhorn2006}, after introducing an external staggered field to induce anti-ferromagnetic correlations in 2D clusters. Putting together the results from this and previous works, the evidence indicates that: (i) these features are not artifacts of the quasi-one-dimensional ladder, (ii) they survive in the presence of long-range order.

\begin{figure}%[ht]
 \includegraphics[width=0.48\textwidth, angle=0]{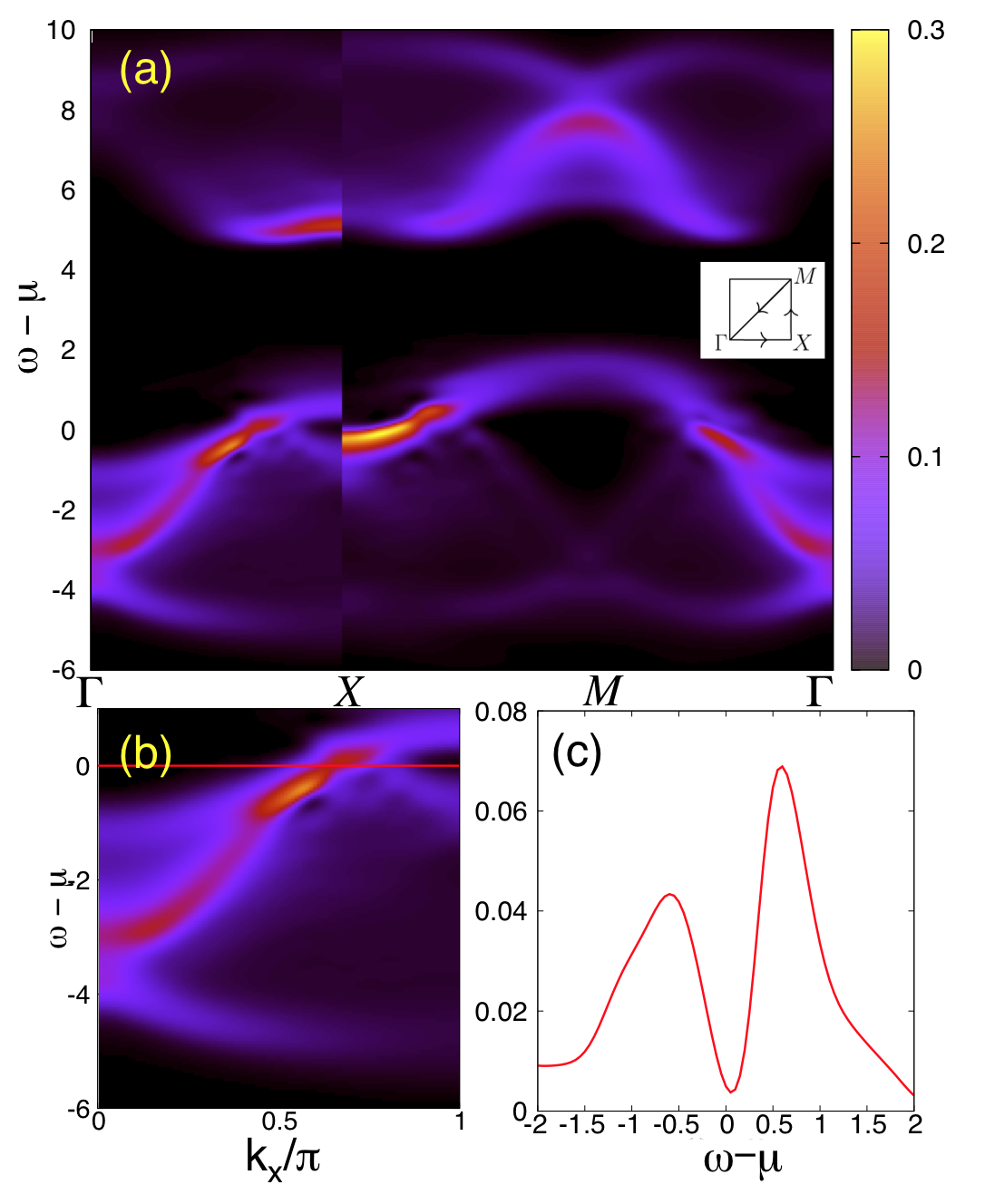}
\caption{
(a) CPT spectral function of the doped $U=8$ 2D Hubbard model obtained using a $2\times 40$ ladder as a cluster, with $n=72$ electrons. (b) Same results focusing on the kink and the pseudogap region along the $\Gamma \rightarrow X$ line. (c) Pseudogap at the $X$ point.
}
\label{fig:doped}
\end{figure}

 We shift our attention now toward the doped case. In Fig.\ref{fig:doped}(a) we 
show a similar calculation for a $2 \times 40$ ladder with 72 electrons, corresponding to $10\%$ doping, which also keeps us away from any charge-density wave instabilities. We used smaller cluster and more states ($m=1000$), since now the charge sector is gapless and introduces more entanglement in the problem, making the simulations computationally more expensive.

The spectrum looks very similar to the CPT results in small clusters \cite{Kohno2012}: the waterfall is no longer a discontinuity but a continuous feature resembling a ``kink'', and there is clear transfer of spectral weight above the Fermi energy centered around the $M$ point (Fig.\ref{fig:doped}(c)). This kink is identical to the one obtained with the DCA in Ref.\onlinecite{Macridin2006}. 
 In addition, our results show an additional ``splitting'' of the bands below and above the Fermi energy along the $k_y=0$ line and centered at around the $X$ point. The splitting of the bands is accompanied by an additional kink at the Fermi surface. This kink appears at the onset of a branch of excitations that could be traced back to the upper branch of the spinon-antiholon continuum in the one-dimensional Hubbard model \cite{Kohno2010,Kohno2012}. Remarkably, these features also appear in DCA calculations \cite{Maier2005}, and CPT calculations on $4\times4$ clusters \cite{Kohno2012}, which in principle should not have any ``memory'' of 1D physics and spin-charge separation. The splitting, though is more marked in our results, and can be interpreted as a pseudogap, as we can clearly see in a cut along the frequency axes in Fig.\ref{fig:doped}(c), in agreement with previous observations.

\section{Discussion}

We have presented a study of the spectral properties of the 2D Hubbard model using the DMRG method as a cluster solver for CPT. Our clusters are ``infinite'' (very long) 2-leg ladders, which already contain information about the thermodynamic limit along the leg spatial direction. In addition, it is reasonable to expect that due to the large size of the ladders, charge fluctuations inside the clusters are largely reduced.
Results show a remarkable resolution of the bands and allow us to identify features such as waterfalls, kinks, and pseudogap, of significance in the physics of cuprate superconductors. We relate these aspects to one-dimensional physics that survives, even in the presence of AFM correlations. We point out that these features are also observed in simulations on 2D clusters, and DCA, indicating that they likely are not artifacts of our cluster choice, and despite breaking rotational symmetry.

Therefore, the main question one should ask would be: what is the fate of spin and charge separation in the presence on long-range antiferromagnetic correlations? 
Whether our spectra display genuine aspects of the physics of the 2D Hubbard model cannot be determined with complete certainty from our results since cluster perturbation theory does not account for the presence of long range antiferromagnetic order in two dimensions. 
Ladders are quasi-1D systems and gapped, with a fast decay of the correlations (Hubbard ladders have a spin correlation length of about 4 lattice spaces\cite{Noack1994} for $U=8$ at half-filling). The spin gap, and the correlation length decay quite rapidly upon doping. In 2D, long range AMF order is also expected to be greatly suppressed away from half-filling. The  
remarkable agreement with Monte Carlo\cite{Bulut1994,Preuss1995,Preuss1997,Grober2000}, VCA \cite{Dahnken2004,Aichhorn2006}, and DCA\cite{Macridin2006} on square clusters, indicates that our ladders contain a great deal of information and display features corresponding to the 2D physics of the Hubbard model. In addittion, 2D-AFM long range order exists only at zero temperature, so it is conceivable that the CPT spectrum is a faithful representation of the excitations of the system at finite $T$, after the correlation length reduces to a few lattice spaces, as also suggested by the aforementioned QMC results\cite{Grober2000}. Further studies to elucidate these questions may have to consider the artificial addition of a staggered magnetic field {\it a la} VCA.

\section{Acklowledgements}
The authors thank NSF for support through Grant No. DMR-1339564.

%%%%%%%%%%%%%%%%%%%%%%%%%%%%%%%%%%%%%%%%%%%%%%%%%%%%%
%This new approach will allow us to study multi orbital problems of great relevance to the fields of superconductivity, and ``Mottronics'', in the bulk, overcoming the limitations of conventional DMRG, and exact diagonalization.

%\bibliography{cpt,mott}

\begin{thebibliography}{72}
\expandafter\ifx\csname natexlab\endcsname\relax\def\natexlab#1{#1}\fi
\expandafter\ifx\csname bibnamefont\endcsname\relax
  \def\bibnamefont#1{#1}\fi
\expandafter\ifx\csname bibfnamefont\endcsname\relax
  \def\bibfnamefont#1{#1}\fi
\expandafter\ifx\csname citenamefont\endcsname\relax
  \def\citenamefont#1{#1}\fi
\expandafter\ifx\csname url\endcsname\relax
  \def\url#1{\texttt{#1}}\fi
\expandafter\ifx\csname urlprefix\endcsname\relax\def\urlprefix{URL }\fi
\providecommand{\bibinfo}[2]{#2}
\providecommand{\eprint}[2][]{\url{#2}}

\bibitem[{\citenamefont{Damascelli et~al.}(2003)\citenamefont{Damascelli,
  Hussain, and Shen}}]{Damascelli2003}
\bibinfo{author}{\bibfnamefont{A.}~\bibnamefont{Damascelli}},
  \bibinfo{author}{\bibfnamefont{Z.}~\bibnamefont{Hussain}}, \bibnamefont{and}
  \bibinfo{author}{\bibfnamefont{Z.-X.} \bibnamefont{Shen}},
  \bibinfo{journal}{Rev. Mod. Phys.} \textbf{\bibinfo{volume}{75}},
  \bibinfo{pages}{473} (\bibinfo{year}{2003}).

\bibitem[{\citenamefont{Ronning et~al.}(2005)\citenamefont{Ronning, Shen,
  Armitage, Damascelli, Lu, Shen, Miller, and Kim}}]{Ronning2005}
\bibinfo{author}{\bibfnamefont{F.}~\bibnamefont{Ronning}},
  \bibinfo{author}{\bibfnamefont{K.~M.} \bibnamefont{Shen}},
  \bibinfo{author}{\bibfnamefont{N.~P.} \bibnamefont{Armitage}},
  \bibinfo{author}{\bibfnamefont{A.}~\bibnamefont{Damascelli}},
  \bibinfo{author}{\bibfnamefont{D.~H.} \bibnamefont{Lu}},
  \bibinfo{author}{\bibfnamefont{Z.-X.} \bibnamefont{Shen}},
  \bibinfo{author}{\bibfnamefont{L.~L.} \bibnamefont{Miller}},
  \bibnamefont{and} \bibinfo{author}{\bibfnamefont{C.}~\bibnamefont{Kim}},
  \bibinfo{journal}{Phys. Rev. B} \textbf{\bibinfo{volume}{71}},
  \bibinfo{pages}{094518} (\bibinfo{year}{2005}).

\bibitem[{\citenamefont{Graf et~al.}(2007)\citenamefont{Graf, Gweon, McElroy,
  Zhou, Jozwiak, Rotenberg, Bill, Sasagawa, Eisaki, Uchida et~al.}}]{Graf2007}
\bibinfo{author}{\bibfnamefont{J.}~\bibnamefont{Graf}},
  \bibinfo{author}{\bibfnamefont{G.-H.} \bibnamefont{Gweon}},
  \bibinfo{author}{\bibfnamefont{K.}~\bibnamefont{McElroy}},
  \bibinfo{author}{\bibfnamefont{S.~Y.} \bibnamefont{Zhou}},
  \bibinfo{author}{\bibfnamefont{C.}~\bibnamefont{Jozwiak}},
  \bibinfo{author}{\bibfnamefont{E.}~\bibnamefont{Rotenberg}},
  \bibinfo{author}{\bibfnamefont{A.}~\bibnamefont{Bill}},
  \bibinfo{author}{\bibfnamefont{T.}~\bibnamefont{Sasagawa}},
  \bibinfo{author}{\bibfnamefont{H.}~\bibnamefont{Eisaki}},
  \bibinfo{author}{\bibfnamefont{S.}~\bibnamefont{Uchida}},
  \bibnamefont{et~al.}, \bibinfo{journal}{Phys. Rev. Lett.}
  \textbf{\bibinfo{volume}{98}}, \bibinfo{pages}{67004} (\bibinfo{year}{2007}).

\bibitem[{\citenamefont{Meng et~al.}(2009)\citenamefont{Meng, Liu, Zhang, Zhao,
  Liu, Jia, Mu, Liu, Dong, Zhang et~al.}}]{Meng2009}
\bibinfo{author}{\bibfnamefont{J.}~\bibnamefont{Meng}},
  \bibinfo{author}{\bibfnamefont{G.}~\bibnamefont{Liu}},
  \bibinfo{author}{\bibfnamefont{W.}~\bibnamefont{Zhang}},
  \bibinfo{author}{\bibfnamefont{L.}~\bibnamefont{Zhao}},
  \bibinfo{author}{\bibfnamefont{H.}~\bibnamefont{Liu}},
  \bibinfo{author}{\bibfnamefont{X.}~\bibnamefont{Jia}},
  \bibinfo{author}{\bibfnamefont{D.}~\bibnamefont{Mu}},
  \bibinfo{author}{\bibfnamefont{S.}~\bibnamefont{Liu}},
  \bibinfo{author}{\bibfnamefont{X.}~\bibnamefont{Dong}},
  \bibinfo{author}{\bibfnamefont{J.}~\bibnamefont{Zhang}},
  \bibnamefont{et~al.}, \bibinfo{journal}{Nature}
  \textbf{\bibinfo{volume}{462}}, \bibinfo{pages}{335} (\bibinfo{year}{2009}).

\bibitem[{\citenamefont{Dagotto et~al.}(1991)\citenamefont{Dagotto, Moreo,
  Ortolani, Riera, and Scalapino}}]{Dagotto1991e}
\bibinfo{author}{\bibfnamefont{E.}~\bibnamefont{Dagotto}},
  \bibinfo{author}{\bibfnamefont{A.}~\bibnamefont{Moreo}},
  \bibinfo{author}{\bibfnamefont{F.}~\bibnamefont{Ortolani}},
  \bibinfo{author}{\bibfnamefont{J.}~\bibnamefont{Riera}}, \bibnamefont{and}
  \bibinfo{author}{\bibfnamefont{D.}~\bibnamefont{Scalapino}},
  \bibinfo{journal}{Phys. Rev. Lett.} \textbf{\bibinfo{volume}{67}},
  \bibinfo{pages}{1918} (\bibinfo{year}{1991}).

\bibitem[{\citenamefont{Ohta et~al.}(1992)\citenamefont{Ohta, Tsutsui,
  Koshibae, Shimozato, and Maekawa}}]{Ohta1992}
\bibinfo{author}{\bibfnamefont{Y.}~\bibnamefont{Ohta}},
  \bibinfo{author}{\bibfnamefont{K.}~\bibnamefont{Tsutsui}},
  \bibinfo{author}{\bibfnamefont{W.}~\bibnamefont{Koshibae}},
  \bibinfo{author}{\bibfnamefont{T.}~\bibnamefont{Shimozato}},
  \bibnamefont{and} \bibinfo{author}{\bibfnamefont{S.}~\bibnamefont{Maekawa}},
  \bibinfo{journal}{Phys. Rev. B} \textbf{\bibinfo{volume}{46}},
  \bibinfo{pages}{14022} (\bibinfo{year}{1992}).

\bibitem[{\citenamefont{Dagotto et~al.}(1992)\citenamefont{Dagotto, Ortolani,
  and Scalapino}}]{Dagotto1992}
\bibinfo{author}{\bibfnamefont{E.}~\bibnamefont{Dagotto}},
  \bibinfo{author}{\bibfnamefont{F.}~\bibnamefont{Ortolani}}, \bibnamefont{and}
  \bibinfo{author}{\bibfnamefont{D.}~\bibnamefont{Scalapino}},
  \bibinfo{journal}{Phys. Rev. B} \textbf{\bibinfo{volume}{46}},
  \bibinfo{pages}{3183} (\bibinfo{year}{1992}).

\bibitem[{\citenamefont{Meinders et~al.}(1993)\citenamefont{Meinders, Eskes,
  and Sawatzky}}]{Meinders1993}
\bibinfo{author}{\bibfnamefont{M.~B.~J.} \bibnamefont{Meinders}},
  \bibinfo{author}{\bibfnamefont{H.}~\bibnamefont{Eskes}}, \bibnamefont{and}
  \bibinfo{author}{\bibfnamefont{G.~a.} \bibnamefont{Sawatzky}},
  \bibinfo{journal}{Phys. Rev. B} \textbf{\bibinfo{volume}{48}},
  \bibinfo{pages}{3916} (\bibinfo{year}{1993}).

\bibitem[{\citenamefont{Eskes et~al.}(1994)\citenamefont{Eskes, Meinders, and
  Stephan}}]{Eskes1994}
\bibinfo{author}{\bibfnamefont{H.}~\bibnamefont{Eskes}},
  \bibinfo{author}{\bibfnamefont{M.~B.} \bibnamefont{Meinders}},
  \bibnamefont{and} \bibinfo{author}{\bibfnamefont{W.}~\bibnamefont{Stephan}},
  \bibinfo{journal}{Phys. Rev. B} \textbf{\bibinfo{volume}{50}},
  \bibinfo{pages}{17980} (\bibinfo{year}{1994}).

\bibitem[{\citenamefont{Preuss et~al.}(1995{\natexlab{a}})\citenamefont{Preuss,
  von~der Linden, and Hanke}}]{Preuss1994a}
\bibinfo{author}{\bibfnamefont{R.}~\bibnamefont{Preuss}},
  \bibinfo{author}{\bibfnamefont{W.}~\bibnamefont{von~der Linden}},
  \bibnamefont{and} \bibinfo{author}{\bibfnamefont{W.}~\bibnamefont{Hanke}},
  \bibinfo{journal}{Phys. Rev. Lett.} \textbf{\bibinfo{volume}{75}},
  \bibinfo{pages}{1344} (\bibinfo{year}{1995}{\natexlab{a}}).

\bibitem[{\citenamefont{Bulut et~al.}(1994)\citenamefont{Bulut, Scalapino, and
  White}}]{Bulut1994}
\bibinfo{author}{\bibfnamefont{N.}~\bibnamefont{Bulut}},
  \bibinfo{author}{\bibfnamefont{D.~J.} \bibnamefont{Scalapino}},
  \bibnamefont{and} \bibinfo{author}{\bibfnamefont{S.~R.} \bibnamefont{White}},
  \bibinfo{journal}{Phys. Rev. Lett.} \textbf{\bibinfo{volume}{72}},
  \bibinfo{pages}{705} (\bibinfo{year}{1994}).

\bibitem[{\citenamefont{Preuss et~al.}(1995{\natexlab{b}})\citenamefont{Preuss,
  Hanke, and von~der Linden}}]{Preuss1995}
\bibinfo{author}{\bibfnamefont{R.}~\bibnamefont{Preuss}},
  \bibinfo{author}{\bibfnamefont{W.}~\bibnamefont{Hanke}}, \bibnamefont{and}
  \bibinfo{author}{\bibfnamefont{W.}~\bibnamefont{von~der Linden}},
  \bibinfo{journal}{Phys. Rev. Lett.} \textbf{\bibinfo{volume}{75}},
  \bibinfo{pages}{1344} (\bibinfo{year}{1995}{\natexlab{b}}).

\bibitem[{\citenamefont{Moreo et~al.}(1995)\citenamefont{Moreo, Haas, Sandvik,
  and Dagotto}}]{Moreo1995c}
\bibinfo{author}{\bibfnamefont{A.}~\bibnamefont{Moreo}},
  \bibinfo{author}{\bibfnamefont{S.}~\bibnamefont{Haas}},
  \bibinfo{author}{\bibfnamefont{A.~W.} \bibnamefont{Sandvik}},
  \bibnamefont{and} \bibinfo{author}{\bibfnamefont{E.}~\bibnamefont{Dagotto}},
  \bibinfo{journal}{Phys. Rev. B} \textbf{\bibinfo{volume}{51}},
  \bibinfo{pages}{12045} (\bibinfo{year}{1995}).

\bibitem[{\citenamefont{Georges et~al.}(1996)\citenamefont{Georges, Kotliar,
  Krauth, and Rozenberg}}]{Georges1996}
\bibinfo{author}{\bibfnamefont{A.}~\bibnamefont{Georges}},
  \bibinfo{author}{\bibfnamefont{G.}~\bibnamefont{Kotliar}},
  \bibinfo{author}{\bibfnamefont{W.}~\bibnamefont{Krauth}}, \bibnamefont{and}
  \bibinfo{author}{\bibfnamefont{M.~J.} \bibnamefont{Rozenberg}},
  \bibinfo{journal}{Rev. Mod. Phys.} \textbf{\bibinfo{volume}{68}},
  \bibinfo{pages}{13} (\bibinfo{year}{1996}).

\bibitem[{\citenamefont{Preuss et~al.}(1997)\citenamefont{Preuss, Hanke,
  Gr{\"{o}}ber, and Evertz}}]{Preuss1997}
\bibinfo{author}{\bibfnamefont{R.}~\bibnamefont{Preuss}},
  \bibinfo{author}{\bibfnamefont{W.}~\bibnamefont{Hanke}},
  \bibinfo{author}{\bibfnamefont{C.}~\bibnamefont{Gr{\"{o}}ber}},
  \bibnamefont{and} \bibinfo{author}{\bibfnamefont{H.~G.}
  \bibnamefont{Evertz}}, \bibinfo{journal}{Phys. Rev. Lett.}
  \textbf{\bibinfo{volume}{79}}, \bibinfo{pages}{1122} (\bibinfo{year}{1997}).

\bibitem[{\citenamefont{Groeber et~al.}(2000)\citenamefont{Groeber, Eder, and
  Hanke}}]{Groeber2000}
\bibinfo{author}{\bibfnamefont{C.}~\bibnamefont{Groeber}},
  \bibinfo{author}{\bibfnamefont{R.}~\bibnamefont{Eder}}, \bibnamefont{and}
  \bibinfo{author}{\bibfnamefont{W.}~\bibnamefont{Hanke}},
  \textbf{\bibinfo{volume}{62}}, \bibinfo{pages}{4336} (\bibinfo{year}{2000}).

\bibitem[{\citenamefont{Huscroft et~al.}(2001)\citenamefont{Huscroft, Jarrell,
  Maier, Moukouri, and Tahvildarzadeh}}]{Huscroft2001}
\bibinfo{author}{\bibfnamefont{C.}~\bibnamefont{Huscroft}},
  \bibinfo{author}{\bibfnamefont{M.}~\bibnamefont{Jarrell}},
  \bibinfo{author}{\bibfnamefont{T.}~\bibnamefont{Maier}},
  \bibinfo{author}{\bibfnamefont{S.}~\bibnamefont{Moukouri}}, \bibnamefont{and}
  \bibinfo{author}{\bibfnamefont{A.~N.} \bibnamefont{Tahvildarzadeh}},
  \bibinfo{journal}{Phys. Rev. Lett.} \textbf{\bibinfo{volume}{86}},
  \bibinfo{pages}{139} (\bibinfo{year}{2001}).

\bibitem[{\citenamefont{S\'{e}n\'{e}chal and Perez}(2000)}]{Senechal2000}
\bibinfo{author}{\bibfnamefont{D.}~\bibnamefont{S\'{e}n\'{e}chal}}
  \bibnamefont{and} \bibinfo{author}{\bibfnamefont{D.}~\bibnamefont{Perez}},
  \bibinfo{journal}{Phys Rev Lett} \textbf{\bibinfo{volume}{84}},
  \bibinfo{pages}{522} (\bibinfo{year}{2000}).

\bibitem[{\citenamefont{Maier et~al.}(2002)\citenamefont{Maier, Pruschke, and
  Jarrell}}]{Maier2002a}
\bibinfo{author}{\bibfnamefont{T.~A.} \bibnamefont{Maier}},
  \bibinfo{author}{\bibfnamefont{T.}~\bibnamefont{Pruschke}}, \bibnamefont{and}
  \bibinfo{author}{\bibfnamefont{M.}~\bibnamefont{Jarrell}},
  \bibinfo{journal}{Phys. Rev. B} \textbf{\bibinfo{volume}{66}},
  \bibinfo{pages}{75102} (\bibinfo{year}{2002}).

\bibitem[{\citenamefont{S{\'{e}}n{\'{e}}chal
  et~al.}(2002)\citenamefont{S{\'{e}}n{\'{e}}chal, Perez, and
  Plouffe}}]{Senechal2002}
\bibinfo{author}{\bibfnamefont{D.}~\bibnamefont{S{\'{e}}n{\'{e}}chal}},
  \bibinfo{author}{\bibfnamefont{D.}~\bibnamefont{Perez}}, \bibnamefont{and}
  \bibinfo{author}{\bibfnamefont{D.}~\bibnamefont{Plouffe}},
  \bibinfo{journal}{Phys. Rev. B} \textbf{\bibinfo{volume}{66}},
  \bibinfo{pages}{075129} (\bibinfo{year}{2002}).

\bibitem[{\citenamefont{Stanescu and Phillips}(2003)}]{Stanescu2003}
\bibinfo{author}{\bibfnamefont{T.~D.} \bibnamefont{Stanescu}} \bibnamefont{and}
  \bibinfo{author}{\bibfnamefont{P.}~\bibnamefont{Phillips}},
  \bibinfo{journal}{Phys. Rev. Lett.} \textbf{\bibinfo{volume}{91}},
  \bibinfo{pages}{017002} (\bibinfo{year}{2003}).

\bibitem[{\citenamefont{Dahnken et~al.}(2004)\citenamefont{Dahnken, Aichhorn,
  Hanke, Arrigoni, and Potthoff}}]{Dahnken2004}
\bibinfo{author}{\bibfnamefont{C.}~\bibnamefont{Dahnken}},
  \bibinfo{author}{\bibfnamefont{M.}~\bibnamefont{Aichhorn}},
  \bibinfo{author}{\bibfnamefont{W.}~\bibnamefont{Hanke}},
  \bibinfo{author}{\bibfnamefont{E.}~\bibnamefont{Arrigoni}}, \bibnamefont{and}
  \bibinfo{author}{\bibfnamefont{M.}~\bibnamefont{Potthoff}},
  \bibinfo{journal}{Phys. Rev. B} \textbf{\bibinfo{volume}{70}},
  \bibinfo{pages}{245110} (\bibinfo{year}{2004}).

\bibitem[{\citenamefont{Aichhorn and Arrigoni}(2005)}]{Aichhorn2005}
\bibinfo{author}{\bibfnamefont{M.}~\bibnamefont{Aichhorn}} \bibnamefont{and}
  \bibinfo{author}{\bibfnamefont{E.}~\bibnamefont{Arrigoni}},
  \bibinfo{journal}{EPL (Europhysics Letters)} \textbf{\bibinfo{volume}{72}},
  \bibinfo{pages}{117} (\bibinfo{year}{2005}).

\bibitem[{\citenamefont{Stanescu et~al.}(2006)\citenamefont{Stanescu, Civelli,
  Haule, and Kotliar}}]{Stanescu2006}
\bibinfo{author}{\bibfnamefont{T.~D.} \bibnamefont{Stanescu}},
  \bibinfo{author}{\bibfnamefont{M.}~\bibnamefont{Civelli}},
  \bibinfo{author}{\bibfnamefont{K.}~\bibnamefont{Haule}}, \bibnamefont{and}
  \bibinfo{author}{\bibfnamefont{G.}~\bibnamefont{Kotliar}},
  \bibinfo{journal}{Ann. Phys.} \textbf{\bibinfo{volume}{321}},
  \bibinfo{pages}{1682} (\bibinfo{year}{2006}).

\bibitem[{\citenamefont{Aichhorn et~al.}(2006)\citenamefont{Aichhorn, Arrigoni,
  Potthoff, and Hanke}}]{Aichhorn2006}
\bibinfo{author}{\bibfnamefont{M.}~\bibnamefont{Aichhorn}},
  \bibinfo{author}{\bibfnamefont{E.}~\bibnamefont{Arrigoni}},
  \bibinfo{author}{\bibfnamefont{M.}~\bibnamefont{Potthoff}}, \bibnamefont{and}
  \bibinfo{author}{\bibfnamefont{W.}~\bibnamefont{Hanke}},
  \bibinfo{journal}{Phys. Rev. B} \textbf{\bibinfo{volume}{74}},
  \bibinfo{pages}{1} (\bibinfo{year}{2006}).

\bibitem[{\citenamefont{Macridin et~al.}(2006)\citenamefont{Macridin, Jarrell,
  Maier, Kent, and D'azevedo}}]{Macridin2006}
\bibinfo{author}{\bibfnamefont{A.}~\bibnamefont{Macridin}},
  \bibinfo{author}{\bibfnamefont{M.}~\bibnamefont{Jarrell}},
  \bibinfo{author}{\bibfnamefont{T.}~\bibnamefont{Maier}},
  \bibinfo{author}{\bibfnamefont{P.~R.~C.} \bibnamefont{Kent}},
  \bibnamefont{and}
  \bibinfo{author}{\bibfnamefont{E.}~\bibnamefont{D'azevedo}},
  \bibinfo{journal}{Phys. Rev. Lett.} \textbf{\bibinfo{volume}{97}},
  \bibinfo{pages}{36401} (\bibinfo{year}{2006}).

\bibitem[{\citenamefont{Tremblay et~al.}(2006)\citenamefont{Tremblay, Kyung,
  and S{\'{e}}n{\'{e}}chal}}]{Tremblay2006}
\bibinfo{author}{\bibfnamefont{A.~S.} \bibnamefont{Tremblay}},
  \bibinfo{author}{\bibfnamefont{B.}~\bibnamefont{Kyung}}, \bibnamefont{and}
  \bibinfo{author}{\bibfnamefont{D.}~\bibnamefont{S{\'{e}}n{\'{e}}chal}},
  \bibinfo{journal}{Low Temp. Phys.} \textbf{\bibinfo{volume}{32}},
  \bibinfo{pages}{424} (\bibinfo{year}{2006}).

\bibitem[{\citenamefont{Kyung et~al.}(2006)\citenamefont{Kyung, Kancharla,
  Senechal, Tremblay, Civelli, and Kotliar}}]{Kyung2006}
\bibinfo{author}{\bibfnamefont{B.}~\bibnamefont{Kyung}},
  \bibinfo{author}{\bibfnamefont{S.~S.} \bibnamefont{Kancharla}},
  \bibinfo{author}{\bibfnamefont{D.}~\bibnamefont{Senechal}},
  \bibinfo{author}{\bibfnamefont{A.~M.~S.} \bibnamefont{Tremblay}},
  \bibinfo{author}{\bibfnamefont{M.}~\bibnamefont{Civelli}}, \bibnamefont{and}
  \bibinfo{author}{\bibfnamefont{G.}~\bibnamefont{Kotliar}},
  \bibinfo{journal}{Phys. Rev. B} \textbf{\bibinfo{volume}{73}},
  \bibinfo{pages}{165144} (\bibinfo{year}{2006}).

\bibitem[{\citenamefont{Stanescu and Kotliar}(2006)}]{Stanescu2006a}
\bibinfo{author}{\bibfnamefont{T.~D.} \bibnamefont{Stanescu}} \bibnamefont{and}
  \bibinfo{author}{\bibfnamefont{G.}~\bibnamefont{Kotliar}},
  \bibinfo{journal}{Phys. Rev. B} \textbf{\bibinfo{volume}{74}},
  \bibinfo{pages}{125110} (\bibinfo{year}{2006}).

\bibitem[{\citenamefont{Sakai et~al.}(2009)\citenamefont{Sakai, Motome, and
  Imada}}]{Sakai2009}
\bibinfo{author}{\bibfnamefont{S.}~\bibnamefont{Sakai}},
  \bibinfo{author}{\bibfnamefont{Y.}~\bibnamefont{Motome}}, \bibnamefont{and}
  \bibinfo{author}{\bibfnamefont{M.}~\bibnamefont{Imada}},
  \bibinfo{journal}{Phys. Rev. Lett.} \textbf{\bibinfo{volume}{102}},
  \bibinfo{pages}{056404} (\bibinfo{year}{2009}).

\bibitem[{\citenamefont{Liebsch}(2010)}]{Liebsch2010}
\bibinfo{author}{\bibfnamefont{A.}~\bibnamefont{Liebsch}},
  \bibinfo{journal}{Phys. Rev. B} \textbf{\bibinfo{volume}{81}},
  \bibinfo{pages}{235133} (\bibinfo{year}{2010}).

\bibitem[{\citenamefont{Sakai et~al.}(2010)\citenamefont{Sakai, Motome, and
  Imada}}]{Sakai2010}
\bibinfo{author}{\bibfnamefont{S.}~\bibnamefont{Sakai}},
  \bibinfo{author}{\bibfnamefont{Y.}~\bibnamefont{Motome}}, \bibnamefont{and}
  \bibinfo{author}{\bibfnamefont{M.}~\bibnamefont{Imada}},
  \bibinfo{journal}{Phys. Rev. B - Condens. Matter Mater. Phys.}
  \textbf{\bibinfo{volume}{82}}, \bibinfo{pages}{1} (\bibinfo{year}{2010}).

\bibitem[{\citenamefont{Kohno}(2010)}]{Kohno2010}
\bibinfo{author}{\bibfnamefont{M.}~\bibnamefont{Kohno}},
  \bibinfo{journal}{Phys. Rev. Lett.} \textbf{\bibinfo{volume}{105}},
  \bibinfo{pages}{106402} (\bibinfo{year}{2010}).

\bibitem[{\citenamefont{Eder et~al.}(2011)\citenamefont{Eder, Seki, and
  Ohta}}]{Eder2011}
\bibinfo{author}{\bibfnamefont{R.}~\bibnamefont{Eder}},
  \bibinfo{author}{\bibfnamefont{K.}~\bibnamefont{Seki}}, \bibnamefont{and}
  \bibinfo{author}{\bibfnamefont{Y.}~\bibnamefont{Ohta}},
  \bibinfo{journal}{Phys. Rev. B} \textbf{\bibinfo{volume}{83}},
  \bibinfo{pages}{205137} (\bibinfo{year}{2011}).

\bibitem[{\citenamefont{Kohno}(2012)}]{Kohno2012}
\bibinfo{author}{\bibfnamefont{M.}~\bibnamefont{Kohno}},
  \bibinfo{journal}{Phys. Rev. Lett.} \textbf{\bibinfo{volume}{108}},
  \bibinfo{pages}{076401} (\bibinfo{year}{2012}).

\bibitem[{\citenamefont{Wang et~al.}(2015)\citenamefont{Wang, Wohlfeld, Moritz,
  Jia, Wu, Chen, and Devereaux}}]{Wang2015a}
\bibinfo{author}{\bibfnamefont{Y.}~\bibnamefont{Wang}},
  \bibinfo{author}{\bibfnamefont{K.}~\bibnamefont{Wohlfeld}},
  \bibinfo{author}{\bibfnamefont{B.}~\bibnamefont{Moritz}},
  \bibinfo{author}{\bibfnamefont{C.~J.} \bibnamefont{Jia}},
  \bibinfo{author}{\bibfnamefont{K.}~\bibnamefont{Wu}},
  \bibinfo{author}{\bibfnamefont{C.}~\bibnamefont{Chen}}, \bibnamefont{and}
  \bibinfo{author}{\bibfnamefont{T.~P.} \bibnamefont{Devereaux}},
  \bibinfo{journal}{Phys. Rev. B} \textbf{\bibinfo{volume}{92}},
  \bibinfo{pages}{075119} (\bibinfo{year}{2015}).

\bibitem[{\citenamefont{Kung et~al.}(2015)\citenamefont{Kung, Nowadnick, Jia,
  Johnston, Moritz, Scalettar, and Devereaux}}]{Kung2015}
\bibinfo{author}{\bibfnamefont{Y.~F.} \bibnamefont{Kung}},
  \bibinfo{author}{\bibfnamefont{E.~A.} \bibnamefont{Nowadnick}},
  \bibinfo{author}{\bibfnamefont{C.~J.} \bibnamefont{Jia}},
  \bibinfo{author}{\bibfnamefont{S.}~\bibnamefont{Johnston}},
  \bibinfo{author}{\bibfnamefont{B.}~\bibnamefont{Moritz}},
  \bibinfo{author}{\bibfnamefont{R.~T.} \bibnamefont{Scalettar}},
  \bibnamefont{and} \bibinfo{author}{\bibfnamefont{T.~P.}
  \bibnamefont{Devereaux}}, \bibinfo{journal}{Phys. Rev. B}
  \textbf{\bibinfo{volume}{92}}, \bibinfo{pages}{195108}
  (\bibinfo{year}{2015}).

\bibitem[{\citenamefont{Eskes et~al.}(1991)\citenamefont{Eskes, Meinders, and
  Sawatzky}}]{Eskes1991}
\bibinfo{author}{\bibfnamefont{H.}~\bibnamefont{Eskes}},
  \bibinfo{author}{\bibfnamefont{M.~B.~J.} \bibnamefont{Meinders}},
  \bibnamefont{and} \bibinfo{author}{\bibfnamefont{G.~A.}
  \bibnamefont{Sawatzky}}, \bibinfo{journal}{Phys. Rev. Lett.}
  \textbf{\bibinfo{volume}{67}}, \bibinfo{pages}{1035} (\bibinfo{year}{1991}).

\bibitem[{\citenamefont{Phillips}(2010)}]{PhillipsRMP2010}
\bibinfo{author}{\bibfnamefont{P.}~\bibnamefont{Phillips}},
  \bibinfo{journal}{Rev. Mod. Phys.} \textbf{\bibinfo{volume}{82}},
  \bibinfo{pages}{1719} (\bibinfo{year}{2010}).

\bibitem[{\citenamefont{Giamarchi}(2004)}]{GiamarchiBook}
\bibinfo{author}{\bibfnamefont{T.}~\bibnamefont{Giamarchi}},
  \emph{\bibinfo{title}{Quantum Physics in One Dimension}}
  (\bibinfo{publisher}{Clarendon Press, Oxford}, \bibinfo{year}{2004}).

\bibitem[{\citenamefont{Gogolin et~al.}(1998)\citenamefont{Gogolin, Nerseyan,
  and Tsvelik}}]{Gogolin}
\bibinfo{author}{\bibfnamefont{A.~O.} \bibnamefont{Gogolin}},
  \bibinfo{author}{\bibfnamefont{A.~A.} \bibnamefont{Nerseyan}},
  \bibnamefont{and} \bibinfo{author}{\bibfnamefont{A.~M.}
  \bibnamefont{Tsvelik}}, \emph{\bibinfo{title}{Bosonization and Strongly
  Correlated Systems}} (\bibinfo{publisher}{Cambridge University Press,
  Cambridge, England}, \bibinfo{year}{1998}).

\bibitem[{\citenamefont{Haldane}(1981)}]{Haldane1981}
\bibinfo{author}{\bibfnamefont{F.~D.~M.} \bibnamefont{Haldane}},
  \bibinfo{journal}{J. Phys. C} \textbf{\bibinfo{volume}{14}},
  \bibinfo{pages}{2585} (\bibinfo{year}{1981}).

\bibitem[{\citenamefont{Anderson}(2000)}]{Anderson2000}
\bibinfo{author}{\bibfnamefont{P.~W.} \bibnamefont{Anderson}},
  \bibinfo{journal}{Phys. C Supercond.} \textbf{\bibinfo{volume}{341-348}},
  \bibinfo{pages}{9} (\bibinfo{year}{2000}).

\bibitem[{\citenamefont{Poilblanc et~al.}(1995)\citenamefont{Poilblanc,
  Scalapino, and Hanke}}]{Poilblanc1995}
\bibinfo{author}{\bibfnamefont{D.}~\bibnamefont{Poilblanc}},
  \bibinfo{author}{\bibfnamefont{D.~J.} \bibnamefont{Scalapino}},
  \bibnamefont{and} \bibinfo{author}{\bibfnamefont{W.}~\bibnamefont{Hanke}},
  \bibinfo{journal}{Phys. Rev. B} \textbf{\bibinfo{volume}{52}},
  \bibinfo{pages}{6796} (\bibinfo{year}{1995}).

\bibitem[{\citenamefont{Haas}(1996)}]{Haas1996a}
\bibinfo{author}{\bibfnamefont{S.}~\bibnamefont{Haas}}, \bibinfo{journal}{Phys.
  Rev. B} \textbf{\bibinfo{volume}{54}}, \bibinfo{pages}{3718}
  (\bibinfo{year}{1996}).

\bibitem[{\citenamefont{Rice et~al.}(1997)\citenamefont{Rice, Haas, Sigrist,
  and Zhang}}]{Rice1997}
\bibinfo{author}{\bibfnamefont{T.}~\bibnamefont{Rice}},
  \bibinfo{author}{\bibfnamefont{S.}~\bibnamefont{Haas}},
  \bibinfo{author}{\bibfnamefont{M.}~\bibnamefont{Sigrist}}, \bibnamefont{and}
  \bibinfo{author}{\bibfnamefont{F.}~\bibnamefont{Zhang}},
  \bibinfo{journal}{Phys. Rev. B} \textbf{\bibinfo{volume}{56}},
  \bibinfo{pages}{14655 } (\bibinfo{year}{1997}).

\bibitem[{\citenamefont{Martins et~al.}(2000)\citenamefont{Martins, Gazza,
  Xavier, Feiguin, and Dagotto}}]{Martins2000}
\bibinfo{author}{\bibfnamefont{G.~B.} \bibnamefont{Martins}},
  \bibinfo{author}{\bibfnamefont{C.}~\bibnamefont{Gazza}},
  \bibinfo{author}{\bibfnamefont{J.~C.} \bibnamefont{Xavier}},
  \bibinfo{author}{\bibfnamefont{A.}~\bibnamefont{Feiguin}}, \bibnamefont{and}
  \bibinfo{author}{\bibfnamefont{E.}~\bibnamefont{Dagotto}},
  \bibinfo{journal}{Phys. Rev. Lett.} \textbf{\bibinfo{volume}{84}},
  \bibinfo{pages}{5844} (\bibinfo{year}{2000}).

\bibitem[{\citenamefont{Brunner et~al.}(2001)\citenamefont{Brunner, Capponi,
  Assaad, and Muramatsu}}]{Brunner2001}
\bibinfo{author}{\bibfnamefont{M.}~\bibnamefont{Brunner}},
  \bibinfo{author}{\bibfnamefont{S.}~\bibnamefont{Capponi}},
  \bibinfo{author}{\bibfnamefont{F.~F.} \bibnamefont{Assaad}},
  \bibnamefont{and}
  \bibinfo{author}{\bibfnamefont{A.}~\bibnamefont{Muramatsu}},
  \bibinfo{journal}{Phys. Rev. B} \textbf{\bibinfo{volume}{63}},
  \bibinfo{pages}{180511} (\bibinfo{year}{2001}).

\bibitem[{\citenamefont{Potthoff}(2003{\natexlab{a}})}]{Potthoff2003}
\bibinfo{author}{\bibfnamefont{M.}~\bibnamefont{Potthoff}},
  \bibinfo{journal}{The European Physical Journal B - Condensed Matter and
  Complex Systems} \textbf{\bibinfo{volume}{36}}, \bibinfo{pages}{335}
  (\bibinfo{year}{2003}{\natexlab{a}}).

\bibitem[{\citenamefont{Potthoff}(2003{\natexlab{b}})}]{Potthoff2003a}
\bibinfo{author}{\bibfnamefont{M.}~\bibnamefont{Potthoff}},
  \bibinfo{journal}{Eur. Phys. J. B - Condens. Matter}
  \textbf{\bibinfo{volume}{32}}, \bibinfo{pages}{429}
  (\bibinfo{year}{2003}{\natexlab{b}}).

\bibitem[{\citenamefont{Potthoff et~al.}(2003)\citenamefont{Potthoff, Aichhorn,
  and Dahnken}}]{Potthoff2003b}
\bibinfo{author}{\bibfnamefont{M.}~\bibnamefont{Potthoff}},
  \bibinfo{author}{\bibfnamefont{M.}~\bibnamefont{Aichhorn}}, \bibnamefont{and}
  \bibinfo{author}{\bibfnamefont{C.}~\bibnamefont{Dahnken}},
  \bibinfo{journal}{Phys. Rev. Lett.} \textbf{\bibinfo{volume}{91}},
  \bibinfo{pages}{206402} (\bibinfo{year}{2003}).

\bibitem[{\citenamefont{Potthoff}(2006)}]{Potthoff2006}
\bibinfo{author}{\bibfnamefont{M.}~\bibnamefont{Potthoff}},
  \bibinfo{journal}{Condens. Matter Phys.} \textbf{\bibinfo{volume}{9}},
  \bibinfo{pages}{557} (\bibinfo{year}{2006}).

\bibitem[{\citenamefont{Kotliar et~al.}(2001)\citenamefont{Kotliar, Savrasov,
  P\'alsson, and Biroli}}]{Kotliar2001}
\bibinfo{author}{\bibfnamefont{G.}~\bibnamefont{Kotliar}},
  \bibinfo{author}{\bibfnamefont{S.~Y.} \bibnamefont{Savrasov}},
  \bibinfo{author}{\bibfnamefont{G.}~\bibnamefont{P\'alsson}},
  \bibnamefont{and} \bibinfo{author}{\bibfnamefont{G.}~\bibnamefont{Biroli}},
  \bibinfo{journal}{Phys. Rev. Lett.} \textbf{\bibinfo{volume}{87}},
  \bibinfo{pages}{186401} (\bibinfo{year}{2001}).

\bibitem[{\citenamefont{Hettler et~al.}(1998)\citenamefont{Hettler,
  Tahvildar-Zadeh, Jarrell, Pruschke, and Krishnamurthy}}]{Hettler1998}
\bibinfo{author}{\bibfnamefont{M.~H.} \bibnamefont{Hettler}},
  \bibinfo{author}{\bibfnamefont{A.~N.} \bibnamefont{Tahvildar-Zadeh}},
  \bibinfo{author}{\bibfnamefont{M.}~\bibnamefont{Jarrell}},
  \bibinfo{author}{\bibfnamefont{T.}~\bibnamefont{Pruschke}}, \bibnamefont{and}
  \bibinfo{author}{\bibfnamefont{H.~R.} \bibnamefont{Krishnamurthy}},
  \bibinfo{journal}{Phys. Rev. B} \textbf{\bibinfo{volume}{58}},
  \bibinfo{pages}{R7475} (\bibinfo{year}{1998}).

\bibitem[{\citenamefont{Hettler et~al.}(2000)\citenamefont{Hettler, Mukherjee,
  Jarrell, and Krishnamurthy}}]{Hettler2000}
\bibinfo{author}{\bibfnamefont{M.~H.} \bibnamefont{Hettler}},
  \bibinfo{author}{\bibfnamefont{M.}~\bibnamefont{Mukherjee}},
  \bibinfo{author}{\bibfnamefont{M.}~\bibnamefont{Jarrell}}, \bibnamefont{and}
  \bibinfo{author}{\bibfnamefont{H.~R.} \bibnamefont{Krishnamurthy}},
  \bibinfo{journal}{Phys. Rev. B} \textbf{\bibinfo{volume}{61}},
  \bibinfo{pages}{12739} (\bibinfo{year}{2000}).

\bibitem[{\citenamefont{Maier et~al.}(2005)\citenamefont{Maier, Jarrell,
  Pruschke, and Hettler}}]{Maier2005}
\bibinfo{author}{\bibfnamefont{T.}~\bibnamefont{Maier}},
  \bibinfo{author}{\bibfnamefont{M.}~\bibnamefont{Jarrell}},
  \bibinfo{author}{\bibfnamefont{T.}~\bibnamefont{Pruschke}}, \bibnamefont{and}
  \bibinfo{author}{\bibfnamefont{M.}~\bibnamefont{Hettler}},
  \bibinfo{journal}{Rev. Mod. Phys.} \textbf{\bibinfo{volume}{77}},
  \bibinfo{pages}{1027 } (\bibinfo{year}{2005}).

\bibitem[{\citenamefont{White and Feiguin}(2004)}]{White2004a}
\bibinfo{author}{\bibfnamefont{S.}~\bibnamefont{White}} \bibnamefont{and}
  \bibinfo{author}{\bibfnamefont{A.}~\bibnamefont{Feiguin}},
  \bibinfo{journal}{Phys. Rev. Lett.} \textbf{\bibinfo{volume}{93}},
  \bibinfo{pages}{076401} (\bibinfo{year}{2004}).

\bibitem[{\citenamefont{Daley et~al.}(2004)\citenamefont{Daley, Kollath,
  Schollw\"ock, , and Vidal}}]{Daley2004}
\bibinfo{author}{\bibfnamefont{A.~J.} \bibnamefont{Daley}},
  \bibinfo{author}{\bibfnamefont{C.}~\bibnamefont{Kollath}},
  \bibinfo{author}{\bibfnamefont{U.}~\bibnamefont{Schollw\"ock}}, ,
  \bibnamefont{and} \bibinfo{author}{\bibfnamefont{G.}~\bibnamefont{Vidal}},
  \bibinfo{journal}{J. Stat. Mech.: Theor. Exp.} p. \bibinfo{pages}{P04005}
  (\bibinfo{year}{2004}).

\bibitem[{\citenamefont{Feiguin and White}(2005)}]{Feiguin2005}
\bibinfo{author}{\bibfnamefont{A.}~\bibnamefont{Feiguin}} \bibnamefont{and}
  \bibinfo{author}{\bibfnamefont{S.}~\bibnamefont{White}},
  \bibinfo{journal}{Phys. Rev. B} \textbf{\bibinfo{volume}{72}},
  \bibinfo{pages}{20404} (\bibinfo{year}{2005}).

\bibitem[{\citenamefont{Feiguin}(2011)}]{vietri}
\bibinfo{author}{\bibfnamefont{A.~E.} \bibnamefont{Feiguin}}, in
  \emph{\bibinfo{booktitle}{XV Training Course in the Physics of Strongly
  Correlated Systems}} (\bibinfo{publisher}{AIP Proceedings},
  \bibinfo{year}{2011}), vol. \bibinfo{volume}{1419}, p.~\bibinfo{pages}{5}.

\bibitem[{\citenamefont{Gross and Valenti}(1993)}]{Gross1993}
\bibinfo{author}{\bibfnamefont{C.}~\bibnamefont{Gross}} \bibnamefont{and}
  \bibinfo{author}{\bibfnamefont{R.}~\bibnamefont{Valenti}},
  \bibinfo{journal}{Phys. Rev. B} \textbf{\bibinfo{volume}{48}},
  \bibinfo{pages}{418} (\bibinfo{year}{1993}).

\bibitem[{\citenamefont{Pereira et~al.}(2009)\citenamefont{Pereira, White, and
  Affleck}}]{Pereira2009}
\bibinfo{author}{\bibfnamefont{R.~G.} \bibnamefont{Pereira}},
  \bibinfo{author}{\bibfnamefont{S.~R.} \bibnamefont{White}}, \bibnamefont{and}
  \bibinfo{author}{\bibfnamefont{I.}~\bibnamefont{Affleck}},
  \bibinfo{journal}{Phys. Rev. B} \textbf{\bibinfo{volume}{79}},
  \bibinfo{pages}{165113} (\bibinfo{year}{2009}).

\bibitem[{\citenamefont{White and Affleck}(2008)}]{White2008}
\bibinfo{author}{\bibfnamefont{S.~R.} \bibnamefont{White}} \bibnamefont{and}
  \bibinfo{author}{\bibfnamefont{I.}~\bibnamefont{Affleck}},
  \bibinfo{journal}{Phys. Rev. B} \textbf{\bibinfo{volume}{77}},
  \bibinfo{pages}{134437} (\bibinfo{year}{2008}).

\bibitem[{\citenamefont{Hallberg}(1995)}]{Hallberg1995}
\bibinfo{author}{\bibfnamefont{K.}~\bibnamefont{Hallberg}},
  \bibinfo{journal}{Phys. Rev. B} \textbf{\bibinfo{volume}{52}},
  \bibinfo{pages}{R9827} (\bibinfo{year}{1995}).

\bibitem[{\citenamefont{K\"uhner and White}(1999)}]{Kuhner1999}
\bibinfo{author}{\bibfnamefont{T.~D.} \bibnamefont{K\"uhner}} \bibnamefont{and}
  \bibinfo{author}{\bibfnamefont{S.~R.} \bibnamefont{White}},
  \bibinfo{journal}{Phys. Rev. B} \textbf{\bibinfo{volume}{60}},
  \bibinfo{pages}{335} (\bibinfo{year}{1999}).

\bibitem[{\citenamefont{Jeckelmann}(2002)}]{Jeckelmann2002}
\bibinfo{author}{\bibfnamefont{E.}~\bibnamefont{Jeckelmann}},
  \bibinfo{journal}{Phys. Rev. B} \textbf{\bibinfo{volume}{66}},
  \bibinfo{pages}{045114} (\bibinfo{year}{2002}).

\bibitem[{\citenamefont{Kohno}(2015)}]{Kohno2015}
\bibinfo{author}{\bibfnamefont{M.}~\bibnamefont{Kohno}},
  \bibinfo{journal}{Phys. Rev. B} \textbf{\bibinfo{volume}{92}},
  \bibinfo{pages}{085129} (\bibinfo{year}{2015}).

\bibitem[{\citenamefont{Luther and Emery}(1974)}]{Luther1974}
\bibinfo{author}{\bibfnamefont{A.}~\bibnamefont{Luther}} \bibnamefont{and}
  \bibinfo{author}{\bibfnamefont{V.~J.} \bibnamefont{Emery}},
  \bibinfo{journal}{Phys. Rev. Lett.} \textbf{\bibinfo{volume}{33}},
  \bibinfo{pages}{589} (\bibinfo{year}{1974}).

\bibitem[{\citenamefont{Balents and Fisher}(1996)}]{Balents1996}
\bibinfo{author}{\bibfnamefont{L.}~\bibnamefont{Balents}} \bibnamefont{and}
  \bibinfo{author}{\bibfnamefont{M.~P.~A.} \bibnamefont{Fisher}},
  \bibinfo{journal}{Phys. Rev. B} \textbf{\bibinfo{volume}{53}},
  \bibinfo{pages}{12133} (\bibinfo{year}{1996}).

\bibitem[{\citenamefont{Gr\"ober et~al.}(2000)\citenamefont{Gr\"ober, Eder, and
  Hanke}}]{Grober2000}
\bibinfo{author}{\bibfnamefont{C.}~\bibnamefont{Gr\"ober}},
  \bibinfo{author}{\bibfnamefont{R.}~\bibnamefont{Eder}}, \bibnamefont{and}
  \bibinfo{author}{\bibfnamefont{W.}~\bibnamefont{Hanke}},
  \bibinfo{journal}{Phys. Rev. B} \textbf{\bibinfo{volume}{62}},
  \bibinfo{pages}{4336} (\bibinfo{year}{2000}).

\bibitem[{\citenamefont{Eder et~al.}(2000)\citenamefont{Eder, Dorneich, Zacher,
  and Gro}}]{Eder2000}
\bibinfo{author}{\bibfnamefont{R.}~\bibnamefont{Eder}},
  \bibinfo{author}{\bibfnamefont{A.}~\bibnamefont{Dorneich}},
  \bibinfo{author}{\bibfnamefont{M.~G.} \bibnamefont{Zacher}},
  \bibnamefont{and} \bibinfo{author}{\bibfnamefont{C.}~\bibnamefont{Gro}},
  \bibinfo{journal}{Phys. Rev. B} \textbf{\bibinfo{volume}{61}},
  \bibinfo{pages}{12816} (\bibinfo{year}{2000}).

\bibitem[{\citenamefont{Noack et~al.}(1994)\citenamefont{Noack, White, and
  Scalapino}}]{Noack1994}
\bibinfo{author}{\bibfnamefont{R.~M.} \bibnamefont{Noack}},
  \bibinfo{author}{\bibfnamefont{S.~R.} \bibnamefont{White}}, \bibnamefont{and}
  \bibinfo{author}{\bibfnamefont{D.~J.} \bibnamefont{Scalapino}},
  \bibinfo{journal}{Phys. Rev. Lett.} \textbf{\bibinfo{volume}{73}},
  \bibinfo{pages}{882} (\bibinfo{year}{1994}).

\end{thebibliography}

\end{document}